\documentclass[10pt,aps,prb,superscriptaddress,twocolumn,citeautoscript,longbibliography]{revtex4-2}
\usepackage[american]{babel}
\usepackage{graphicx}
\usepackage{amsfonts}
\usepackage{amsmath}
\usepackage{amssymb}
\usepackage{bm}
\usepackage{textcomp}
\usepackage{verbatim}
\usepackage{xspace}
\usepackage{soul}
\usepackage{bbold}
\usepackage{mathtools}
\usepackage{dcolumn}
\usepackage{hyperref}
\usepackage[usenames]{color}
\usepackage{subfig}
\usepackage{physics}
\usepackage{units}
\usepackage[export]{adjustbox}
\usepackage{tablefootnote}
\usepackage[dvipsnames]{xcolor}
\usepackage{ragged2e}
\DeclareCaptionJustification{justified}{\justifying}
\def\beq{\begin{equation}}
	\def\eeq{\end{equation}}
\def\vk{{\bf k}}

\def\vR{{\bf R}}

\def\vq{{\bf q}}

\newcommand{\newver}[1]{\textcolor[rgb]{0.0, 0.0, 0.0}{#1}}

\newcommand{\out}[1]{{}}

\begin{document}
	
\title{5$d$-mediated indirect exchange and effective spin Hamiltonians in Ce triangular-lattice delafossites}

	\author{Leonid V. Pourovskii}    \email{leonid.poyurovskiy@polytechnique.edu}
	\affiliation{CPHT, CNRS, \'Ecole polytechnique, Institut Polytechnique de Paris, 91120 Palaiseau, France}
	\affiliation{Coll\`ege de France, Université PSL, 11 place Marcelin Berthelot, 75005 Paris, France}

	\begin{abstract}
		Anisotropic intersite exchange interactions in frustrated rare-earth magnets are difficult to assess both theoretically and experimentally. Here, we propose an ab initio force-theorem framework combining the quasi-atomic Hubbard-I approach to 4$f$ correlations  with a static mean-field  treatment of the on-site intershell Coulomb interaction between rare-earth 4$f$ and  5$d$ states to simultaneously capture  both 4$f$ superexchange and $5d$-mediated indirect exchange. Applying it to the triangular lattice Ce delafossites CsCeSe$_2$, KCeS$_2$, and RbCeO$_2$, we find that the indirect exchange dominates in the selenide, the superexchange in the oxide, while both mechanisms contribute almost equally in the sulfide.  The magnetic exciation spectra of CsCeSe$_2$ and KCeS$_2$ evaluated from the calculated spin Hamiltonains are in good qualitative and quantitative agreement with experimental data. 

	\end{abstract}
	
	
\maketitle
{\it Introduction.} 
Rare-earth magnetism stems from the partually filled rare-earth (RE) 4$f$ shells, where a strong spin-orbit coupling, weaker crystal-field effects and anisotropic intersite exchange interactions (IEI) intertwine to induce an enormous variety of magnetic phases. Those include a rich zoo of magnetic orders like, for example, incommensurate helical orders in heavy RE metals~\cite{Jensen1991}, ferromagnetic RE  semiconductors~\cite{Mauger1986}, primary multipolar orders in the Ce hexaboride~\cite{Shiina1997,Cameron2016}, PrO$_2$~\cite{Santini2009,Khmelevskyi2024}, Pr "1-2-20" cage compounds~\cite{Onimaru2016}, as well as even more exotic quantum spin  liquid phases. The latter are suspected to occur in RE pyrochlores~\cite{Gingras2014,Sibille2017,Smith2025} and in RE triangular-lattice  lattice systems like the delafossites $ARX_2$ (where $R$ is  a 3+ rare-earth ion, $X$ is a ligand and $A$ is a non-magnetic cation)~\cite{Xie2024} or the heptatantalates $R$Ta$_7$O$_{19}$~\cite{Arh2022}. 

Apart from few possible exceptions like $\alpha$-Ce~\cite{Johansson1974},  direct hopping between 4$f$ shells is believed to be negligible. IEI thus arise due to indirect 4$f$ hopping through ligand states leading to superexchange, or due to indirect exchange involving the on-site Coulomb interaction between 4$f$ and 5$d$ states. Both mechanism are believed to contribute comparably in the Eu chalcogenides~\cite{Kasuya1970,Mauger1986}, where the Eu$^{2+}$ ion possesses no orbital moment,
 simplifying the analysis of kinetic exchange processes. A more general case of both spin and orbital RE moment being non-zero and forming a Hund's rule ground state  in accordance with the $LS$  coupling scheme is though  far more common. In this case one may expect an essential IEI anisotropy as well as high-rank multipoles  playing an important role in the case of a large crystal-field  degeneracy of the ground state~\cite{Santini2009}.  Few theoretical approaches are currently available to treat this general case~\cite{Pourovskii2016,Iwahara2022,Otsuki2024},  with their range of applicability not clearly understood ~\cite{hidden_order_review}.

The lack of quantitative theoretical approaches to evaluate intersite exchange in realistic RE systems is apparent in the case of RE delafossites $ARX_2$ hosting a perfect triangular lattice of $R$=Ce, Yb ions. A strong crystal field on the  $R$ 4$f$ shell leads to a pseudo-spin-1/2 ground-state Kramers doublet  that is energetically well isolated from exited crystal-field levels. A Dirac quantum spin liquid phase and its derivatives are suggested to occur in the spin-1/2 triangular lattice model for a certain range of IEI~\cite{Zhu2015,Iqbal2016,Hu2015,Zhu2018,Wu2021,Sherman2023,Wietek2024,Pourovskii2025_with_Alex}. Experimentally, strong evidence for possible quantum spin liquid phases was found in several Yb delafossites  $A$Yb$X_2$ ($A$=K, Na and $X$=O, S, Se)~\cite{Bordelon2017,Sarkar2019,Dai2021,Scheie2024}. In contrast, only conventional magnetic orders have been  reported so far in isostructural Ce systems like, e.~g., KCeO$_2$~\cite{Bordelon2021}, KCeS$_2$~\cite{Bastien2020,Kulbakov2021,Avdoshenko2022}, CsCeSe$_2$~\cite{Xie2024prl,Xie2024prb}, with a possible exception of RbCeO$_2$~\cite{Ortiz2022}. 
 The effective spin-1/2 Hamiltonians of these delafossites remain rather uncertain due to the lack of single crystals, high magnetic fields required to achieve full saturation, and the difficulty of unambiguously determining small anisotropic IEI from fitting magnetic excitations~\cite{Xie2024prl,Xie2024}. Nevertheless, for several systems such experimental fits have been carried out~\cite{Kulbakov2021,Avdoshenko2022,Scheie2024,Xie2024prl}, in particular, for KCeS$_2$~\cite{Kulbakov2021,Avdoshenko2022} and CsCeSe$_2$~\cite{Xie2024prl,Xie2024prb} that exhibit a simple quasi-collinear ($yz$-)stripe antiferromagnetic  order. 
 
 In spite of great interest in these systems, there have been only few attempts to elucidate their spin Hamiltonians theoretically~\cite{Avdoshenko2022,Villanova2023,Pourovskii2025_with_Alex}. Small nearest-neighbor (NN) anisotropies as well as  next-nearest-neighbor (NNN) couplings with a typical magnitude below one Kelvin~\cite{Xie2024} are hard to reliably extract by direct total-energy density-functional-theory(DFT)-based calculations. The magnitude of NNN superexchange  calculated for Yb delafossites using DFT tight-binding 4$f$ hopping is far too low to account  for  the putative quantum spin liquid state  and drastically  disagrees with experimental estimates~\cite{Villanova2023}. Two-site cluster calculations for KCeS$_2$~\cite{Avdoshenko2022} predict reasonable IEI magnitudes and reproduce its $yz$-stripe order   but do not account for some salient features of the excitation spectra. 
 
Full spin Hamiltonians for a set of Ce delafossites were also calculated~\cite{Pourovskii2025_with_Alex} using the force theorem in Hubbard-I (FT-HI)~\cite{Pourovskii2016}. This method extracts  all kinetic exchange couplings  between localized shells   
from its paramagnetic electronic structure  obtained using DFT+dynamical mean-field theory (DMFT)~\cite{Georges1996,Anisimov1997_1,Lichtenstein_LDApp} in conjunction with the quasi-atomic Hubbard-I (HI) approximation for on-site correlations~\cite{hubbard_1}. \newver{The FT-HI method is aimed at correlated insulators or at localized states coupled through itinerant metallic bands. It evaluates all IEI matrix elements within a chosen ground-state manifold, including all possible high-rank  multipolar couplings allowed by the ground-state degeneracy. The method has been  shown to reliably capture high-rank multipolar orders~\cite{Santini2009,Kuramoto2009,hidden_order_review} in actinide dioxides and transition-metal double perovskites as well as anisotropic dipolar and multipolar interactions in  magnetically ordered systems, 
	see, e.~g., Refs.~\cite{Pourovskii2019,Pourovskii2021f,Pourovskii2021,Khmelevskyi2024,FioreMosca2025} as well as Ref.~\cite{hidden_order_review} for a review.} However, in the case of Ce delafossites, the predicted IEI magnitude, though agreeing with experimental estimates for the oxides systems,  appeared to be severely underestimated for the sulfide KCeS$_2$, hinting at other coupling mechanisms beyond $4f$ kinetic exchange playing a key role in this compound. 

Here, we generalize the FT-HI approach to include all  indirect exchange contributions stemming from the 4$f$-5$d$ on-site  Coulomb repulsion. We first introduce this Coulomb interaction in  its full form; treating it on a static mean-field level, we show that the FT-HI idea -- deriving IEI by considering the DFT+DMFT functional response to simultaneous fluctuations on two correlated shells in the paramagnetic phase -- can be seamlessly extended to the 5$d$-mediated indirect exchange. Applying the generalized FT-HI to a set of Ce delafossites -- the selenide CsCeSe$_2$, sulfide KCeS$_2$, and oxide RbCeO$_2$ -- we find a remarkable qualitative evolution of IEI. While the superexchange is found to dominate over the  indirect exchange in the oxide, the situation is reversed in the selenide. Both coupling mechanisms are found to contribute comparably in the sulfide. The calculated IEI place  CsCeSe$_2$ and KCeS$_2$ in the $yz$-stripe region of the published spin-1/2 triangular-lattice model phase diagram, in agreement with experiment, while those for RbCeO$_2$ correspond to a 120$^{\circ}$ antiferromagnet. We evaluate inelastic neutron scattering spectra for   CsCeSe$_2$ and KCeS$_2$ finding  that the calculated IEI provide a qualitative and quantitative agreement with single-crystal and polycrystalline data~\cite{Xie2024prl,Avdoshenko2022} on a level comparable to  the previously published experimental fits.

{\it Generalized force theorem in Hubbard-I.}  We start by supplementing the one-electron Kohn-Sham Hamiltonian  with on-site intra-shell $ff$ and inter-shell $fd$ Coulomb interactions
\beq\label{eq:HU}
H=\sum_{\vk \nu}\epsilon_{\vk\nu}c^{\dagger}_{\vk\nu}c_{\vk\nu}+\sum_{i\in RE} H_{U,(i)}^{ff} + H_{U,(i)}^{fd} -H_{\mathrm{DC},(i)},
\eeq
where $c^{\dagger}_{\vk\nu}$ creates the Bloch state $\nu$ with energy $\epsilon_{\vk\nu}$ at the $\vk$-point in the Brillouin zone, the second sum is over all RE sites $i$. The four-index on-site $ff$ and $fd$ Coulomb repulsion  $H_{U}^{ff}=1/2\langle ab|U_{ff}|cd\rangle f^{\dagger}_af^{\dagger}_bf_df_c$ and $H_{U}^{fd}=\langle a\alpha|U_{fd}|b\beta\rangle f^{\dagger}_ad^{\dagger}_{\alpha}d_{\beta}f_{b}$ are defined in local $f$ and $d$-shell correlated bases, where we label the $f$ and $d$ orbitals using a combined (Latin and Greek, respectively) spin-orbital index ($a\equiv m_a^f\sigma_a^f$ etc.), summation over repeated indices is implied here and below.  $H_{\mathrm{DC},(i)}$ comprises double-counting correction terms for the 4$f$ and 5$d$ shells.
	
Within the DMFT framework, we use the HI approximation to treat the $ff$ interaction between localized 4$f$ orbitals, while the $H_{U}^{fd}$ term involving dispersive 5$d$ bands  is mean-field decoupled, resulting in a static $d$-shell Hartree-Fork self-energy with matrix elements
 \beq\label{eq:Sig_dd}
 \left[\Sigma^{\mathrm{HF},d}\right]_{\alpha\beta}=\left[\langle a\alpha|U_{fd}|b\beta\rangle - \langle a\alpha|U_{fd}|\beta b\rangle\right] \langle n_{ab}\rangle,
 \eeq
where $n_{ab}=f^{\dagger}_af_b,$  and an analogous term $\Sigma^{\mathrm{HF},f}$ for the $f$-shell.

We assume that in the high-temperature paramagnetic phase modeled by DFT+HI, with the 4$f$ shell localized due to  the $ff$ repulsion, the effect of $H_{U}^{fd}$ is properly included on the static level within DFT.  $\Sigma^{\mathrm{HF}^{d(f)}}$ is thus fully compensated by the corresponding double-counting term in (\ref{eq:HU}). 
\newver{Therefore, we start by charge self-consistent DFT+HI calculations of the target compounds including only the intra-shell $ff$ Coulomb term in order to obtain their high-temperature paramagnetic electronic structure, analogously to previous numerous applications of the DFT+HI framework to RE systems, see, e.~g., Refs.~\cite{Lebegue2005,Pourovskii2009,Shick2009,Locht2016,Delange2017}.
The assumption of not including $U_{fd}$ explicitly at the DFT+HI stage is validated by the quantitatively good description of the crystal-field splitting that we obtain for  all Ce delafossites under consideration (\cite{Pourovskii2025_with_Alex}, see also the Supplemental Material (SM)~\cite{supplmat}).}

\newver{Though the DFT+HI approach can reliably reproduce the paramagnetic electronic structure and crystal-field splitting in RE magnets, it is not able to capture intersite kinetic exchange and the resulting magnetic order, since the hybridization function is neglected while solving the DMFT impurity problem within the HI approximation. The FT-HI approach gets around this limitation by introducing small on-site symmetry-breaking fluctuations (magnetic moments) on two neighboring sites in the  paramagnetic DFT+HI electronic structure  and evaluating the system's response to such fluctuations. } 
Correspondingly, here we focus on the 4$f$-shell ground-state multiplet, labeled by the (pseudo)angular momentum $J$, of dimension $N = 2J + 1$, with its states labeled by the projection $M$, and consider small fluctuations around its paramagnetic configuration.  An on-site $M_1M_2$ fluctuation is  encoded as $\epsilon\rho^{M_1M_2}$,   where $\epsilon$ is a small parameter and $\rho^{M_1M_2}$ is a traceless $N\times N$ density matrix  operator. It elements read $\rho^{M_1M_2}_{MM'}=\left[\delta_{MM_1}\delta_{M'M_2}-\delta_{MM'}I/N\right]$, where $I$ is the unit matrix and the second term in the brackets compensates the change of the trace for diagonal fluctuations. Within the FT-HI, matrix elements of the IEI Hamiltonian within the ground-state-multiplet manifold read 
are obtained from  the response of the DFT+DMFT grand potential $\Omega$~\cite{Kotliar2006,Georges2004} to such fluctuations on two neighboring sites $i$ and $j$:
\beq\label{eq:der}
\langle M^i_1M_3^j|H_{\mathrm{IEI}}|M^i_2M^j_4\rangle=\frac{\delta^2\Omega}{\delta\rho^{M_1M_2}_i\delta\rho^{M_3M_4}_j}.
\eeq
 In the force-theorem spirit~\cite{Mackintosh1980,Liechtenstein1987,Szilva2023} only the kinetic-energy term 
 in $\Omega$ was shown to contribute to (\ref{eq:der}). 
 \newver{This term includes both full hybridization through the Kohn-Sham Hamiltonian and the self-energy (see SM~\cite{supplmat}). The resulting expression for the IEI matrix elements combined the on-site response  $\delta\Sigma_i/\delta \rho_i$ with intersite  electron hopping described by the corresponding intersite  Green's function.}
 
 In the generalized FT-HI, the $\delta\Sigma_i/\delta \rho_i$ matrix consists of two diagonal $ff$ and $dd$ blocks given by $\delta\Sigma^{\mathrm{HI}}_i/\delta \rho_i^{M_1M_2}$ and  $\delta\Sigma^{\mathrm{HF},d}_i/\delta \rho_i^{M_1M_2}$, respectively. The dynamical HI $ff$ block  was derived before~\cite{Pourovskii2016}.  \newver{It was shown to correctly account for the lowest order contribution to kinetic exchange $\sim t^2/U$~\cite{Pourovskii2016}, with beyond-HI corrections to the self-energy due to hybridization effects contributing, apparently, only to higher
orders of kinetic exchange.}

 In order to obtain the $dd$ block we map the many-electron $J$-space density matrix  $\rho^{M_1M_2}$ into the 4$f$ one-electron density matrix $n_{ab}=\mathrm{Tr}[n^J_{ab}\cdot \rho^{M_1M_2}]$, where $\left[n^J_{ab}\right]_{MM'}=\langle JM|f^{\dagger}_af_b|JM'\rangle$ \cite{Fioremosca2022}. Substituting this into (\ref{eq:Sig_dd}) we find 
 \beq\label{eq:dSig_dd}
\delta\Sigma^{\mathrm{HF},d}_{\alpha\beta}/\delta \rho^{M_1M_2}=\mathrm{Tr}\left[U_{fd}^{J,\alpha\beta}\cdot \rho^{M_1M_2}\right],
 \eeq
 where $U_{fd}^{J,\alpha\beta}=\left[\langle a\alpha|U_{fd}|b \beta \rangle-\langle a\alpha|U_{fd}|\beta b\rangle\right] n^J_{ab}$.
 
The subsequent derivation is fully analogous to that of the original FT-HI, see Sec.~IIB in Ref.~\cite{Pourovskii2016}, leading to the expression of the same form
\beq\label{eq:FTHI} \langle M^i_1M_3^j|H_{\mathrm{IEI}}|M^i_2M^j_4\rangle=\mathrm{Tr} \left[G_{ij}\frac{\delta\Sigma_j}{\delta \rho_j^{M_3M_4}}G_{ji}\frac{\delta\Sigma_i}{\delta \rho_i^{M_1M_2}}\right],
\eeq
 where the trace is over the Matsubara frequencies (omitted for brevity), and orbital indices. However,  $\delta\Sigma/\delta \rho$ is now a block-diagonal matrix with its $ff$ and $dd$ blocks defined above. Correspondingly, the one-electron propagator matrix $G_{ij}$ is composed of blocks $[G_{ij}]_{bb'}=\mathrm{FT}_{ij}\left[P_{\vk,b} G_{\vk}P_{\vk,b'}^{\dagger}\right]$, where  $\mathrm{FT}_{ij}$ is the Fourier transform evaluated at the lattice vector  $\vR_{ij}$,  $b(b')=f,d$ are the block labels,  $P_{\vk,b}$ are the projectors to the corresponding correlated basis, $G_{\vk}$ is the DFT+HI lattice Green's function. One sees that the propagators   $[G_{ij}]_{bb'}$ include  effective $ff$ hopping leading to superexchange as well as the $fd$ ($df$) and $dd$ blocks that  induce various indirect exchange contributions.


The generalized FT-HI approach is implemented in the framework of the MagInt code~\cite{magint} using the TRIQS library~\cite{Parcollet2015,Aichhorn2016} and Wien2k linearized augmented-plane-wave (LAPW) DFT code ~\cite{Wien2k}.

\begin{figure}[!tb]
	\centering
	\includegraphics[width=0.9\columnwidth]{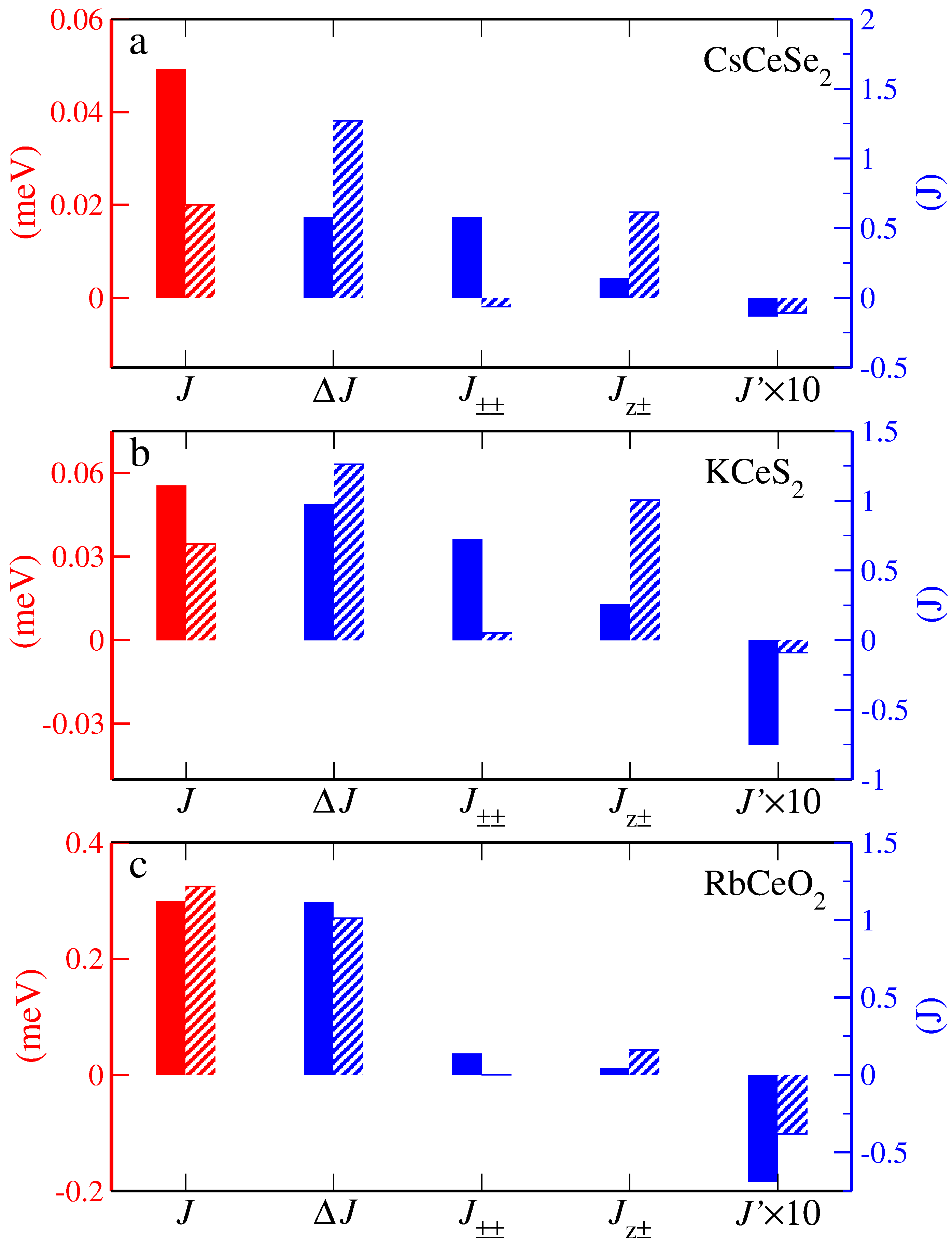}
	\caption{The calculated nearest-neighbor intersite exchange interaction  $J$  as well as the nearest-neighbor exchange anisotropy terms $\Delta J$, $J_{\pm\pm}$, $J_{z\pm}$, and next-nearest-neighbor $J'$  calculated by FT-HI method. \newver{The value of $J$ (red bars)  in meV is indicated by the left $y$-axis that is also colored in red. Other quantities (blue bars) are given in the units of $J$ as indicated by the right $y$-axis of the same color.} 
    The values obtained using the generalized FT-HI approach with the 5$d$-mediated indirect exchange included are shown as solid bars, whereas the hatched bars represent the superexchange-only values. }
	\label{fig:J}
\end{figure}

{\it Calculational parameters.}
First, we carried self-consistent DFT+HI calculations to obtain the paramagnetic electronic structure and ground-state Kramers doublet of the target compounds, and then computed their effective spin Hamiltonians using the generalized FT-HI. \newver{The $fd$ rotationally invariant Coulomb vertex was defined~\cite{Cowan_book}  using the Slater integrals values extracted from optical measurements of $4f^n\to 4f^{n-1}5d^1$ excitations in RE-doped in LiYF$_4$ host~\cite{Burdick2007}. Those $fd$ Slater parameters are
thus assumed to be compound independent. This is an excellent approximation for the intra-$f$-shell
Slater parameters $F^a$ for $a=2,4,6$, which values are known to be virtually completely independent of the
crystalline environment. Its applicability to the $fd$ Slater parameters remains to be assessed. We note that the  $fd$ Slater parameter  $F^0$, which is expected to be strongly system dependent, does not contribute to (\ref{eq:dSig_dd}) since  the fluctuations $\delta \rho^{M_1M_2}$ conserve the 4$f$ shell occupancy. See the SM for other calculational details~\cite{supplmat}. }

\begin{figure}[!t]
	\centering
	\includegraphics[width=1\columnwidth]{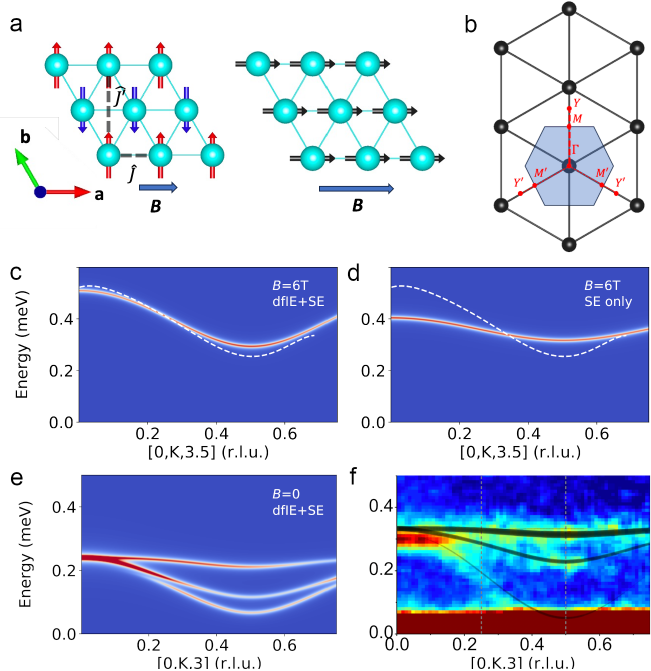}
	\caption{(a). Top view of the  $yz$-stripe order 
    (left) and the fully-polarized state 
    (right). (b) Top view of the reciprocal triangular lattice. The first Brillouin zone is shaded, the $K=\Gamma-M-Y$ path along which the INS in applied field is calculated is indicated by  dashed red line, the thinner lines show the two equivalent $\Gamma-M'-Y'$ paths. (c) The calculated INS intensity along the path $[0,K,3.5]$ in applied field of 6 Tesla using full IEI with 5$d$-mediated indirect exchange (dfIE) included. The experimental dispersion extracted from the experimental INS spectrum~\cite{Xie2024prl} at the same field is shown by the white dashed line.  (d) The same data as in (c) calculated using superexchange (SE) only. (e) The zero-field INS calculated along $[0,K,3]$ using full IEI.  (f). The experimental INS data at zero field~\cite{Xie2024prl}. The black lines were  calculated with their experimental IEI fit.  The figure is reproduced from Ref.~\cite{Xie2024prl} under CC-BY 4.0 license.  }
	\label{fig:CsCeSe2}
\end{figure}

{\it Effective spin Hamiltonians and magnetic order.} 
The delafossites' structure symmetry constrains~\cite{Li2015,Zhu2018,Iaconis2018,Maksimov2019} the most general  IEI within the ground-state spin-1/2 manifold along the NN  $\vR_{ij}$ bond $x||[100]$   (see Fig.~\ref{fig:CsCeSe2}a) to the following form:
\begin{equation}
	\begin{aligned}
		H_{ij}& =\mathbf{S}_i^T \hat{J}_{ij}\mathbf{S}_j=J(\Delta S_i^zS_j^z+S_i^xS_j^x+S_i^yS_j^y)+ \\
		& 2J_{\pm\pm}(S_i^xS_j^x-S_i^yS_j^y) +J_{z\pm}(S_i^zS_j^y+S_i^yS_j^z),
	\end{aligned}
	\label{eq:HNN}
\end{equation}
where $\mathbf{S}=[S^x,S^y,S^z]$, the parameters  $\Delta$, $J_{\pm\pm}$, and $J_{z\pm}$  determine the diagonal XXZ, diagonal XY and off-diagonal exchange anisotropy, respectively.  The coupling $\hat{J'}$ for the NNN bond along $[010]||y$ takes the same form (\ref{eq:HNN}). The IEI for other NN and NNN bonds follow from the symmetry \cite{Iaconis2018,Maksimov2019,Pourovskii2025_with_Alex}. 
%

The key result of this work is shown in Fig.~\ref{fig:J}  where the IEI calculated by the generalized FT-HI are depicted together with the values obtained without the $fd$ contribution, i.~e. including the $f-f$ superexchange  only (the latter, for RbCeO$_2$ and KCeS$_2$,  closely agreeing with previous FT-HI  results~\cite{Pourovskii2025_with_Alex}). The difference between those two IEI sets is due to    5$d$-mediated $f-f$ indirect exchange (dfIE). 
The dfIE contribution dominates in the selenide, whereas  the dfIE and superexchange contribute almost equally in the sulfide. In the oxide, in contrast, the contribution due to superexchange is by far the most significant. Notably, the magnitude of the dfIE term remains comparable across all three systems,  and the observed evolution is due to the magnitude of superexchange being strongly suppressed with decreasing electronegativity and increasing ionic radius along the chalcogen series O$\to$S$\to$Se. The impact of dfIE on the exchange anisotropy and long-distance NNN couplings remains significant even in RbCeO$_2$. It is absolutely crucial in CsCeSe$_2$ and KCeS$_2$ changing the leading NN exchange anisotropy from the off-diagonal $J_{z\pm}$ to the diagonal $XY$ $J_{\pm\pm}$ and, in KCeS$_2$, also  strongly enhancing the NNN coupling (see the SM~\cite{supplmat} for the full list of calculated IEI).

Two physically distinct exchange processes contribute to the dfIE~\cite{Kasuya1970,Mauger1986}. The first one is generated by  intershell $H_U^{fd}$  on two RE sites mediated by $dd$ virtual hopping, while the second one, mediated by  $fd$ hopping, occurs due to  intrashell $H_U^{ff}$  on one RE site and intershell $H_U^{fd}$  on another one. The first process 
involves no  hopping from 4$f$-shell and previously analytically estimated to be negligible in Eu chalcogenides compared to the second one~\cite{Kasuya1970,Mauger1986}. In the generalized FT-HI we can single out the first process by setting the $ff$-block  of $\delta \Sigma/\delta \rho$ to zero. One can easily see that only $dd$ hopping described by the corresponding propagator $[G_{ij}]_{dd}$ contributes in this case. We find that the $dd$ hopping-mediated process contributes overall  about 10\% of the total dfIE magnitude, its major part thus being due to $fd$ virtual hopping.

The ground-state magnetic phases corresponding to the calculated IEI can be inferred from published phase diagrams for the anisotropic triangular-lattice model~\cite{Zhu2018,Maksimov2019,Wu2021,Pourovskii2025_with_Alex}. 
Both CsCeSe$_2$ and KCeS$_2$  lie deep within the region of $yz$-stripe antiferromagnetic order (schematically depicted in Fig.~\ref{fig:CsCeSe2}a), owing to  large diagonal $J_{\pm\pm}$, in agreement with experiment~\cite{Kulbakov2021,Xie2024prb}. The same order is actually stable with only superexchange  included, but due to a different reason, namely, due to the off-diagonal exchange anisotropy $J_{z\pm}$.  
Finally, basing on the exact-diagonalization analysis of Ref.~\cite{Pourovskii2025_with_Alex}, RbCeO$_2$ is expected to be in the 120$^{\circ}$-antiferromagnet region owing to its relatively significant negative $J'$,  in spite of a report~\cite{Ortiz2022} finding no evidence for magnetic order in its thermodynamics.

{\it Magnetic excitations.}
Magnetic excitations  measured with inelastic neutron scattering (INS)  in CsCeSe$_2$ and KCeS$_2$~\cite{Xie2024prl,Avdoshenko2022} provide valuable data to test the calculated spin Hamiltonians. 
In particular, the INS experiment~\cite{Xie2024prl}  in CsCeSe$_2$ was carried out in a single-crystal mode under applied field allowing to resolve excitations along different Brillouin-zone  directions. In small fields the $yz$-stripe phase is preserved, while in high fields of 5~T and above the system is in a fully-polarized  ferromagnetic state (see Fig.~\ref{fig:CsCeSe2}a) featuring sharp magnon excitations. 

In the high-field limit,  magnetic excitations in CsCeSe$_2$ are expected to  be reasonably well captured within simple approaches~\cite{Xie2024prl} like the random-phase approximation (see the SM~\cite{supplmat} for details). 
To calculate  CsCeSe$_2$ magnetic excitation spectra in the random-phase approximation, we first solved its effective Hamiltonian within a single-site mean-field approximation using the McPhase package~\cite{Rotter2004} applying the field along the $a$ direction. 
In zero field we obtained the expected $yz$-stripe order with a small out-of-plane component, $|\langle S^y \rangle|=0.50$ and   $|\langle S^z \rangle|=0.06$. The calculated in-plane ordered moment 0.98$\mu_B$ is significantly overestimated compared to the neutron diffraction value of 0.65$\mu_B$, though agreeing reasonably well with experimental INS estimate (0.88 $\mu_B$)~\cite{Xie2024prb,Xie2024prl}. This suggests strong fluctuations reducing the ordered moment with respect to its "atomic" MF value. 
 
With such effects being beyond our simple mean-field random-phase-approximation framework, we  mainly focus on comparing  calculated excitations  in CsCeSe$_2$ at high fields. 
Since experimentally the Ce local moment appears  to be field-dependent even in the fully-polarized phase~\cite{Xie2024prb}, we compare the measured~\cite{Xie2024prl} and calculated INS data at the applied field of 6 Tesla, at which the experimental value of the in-plain moment~\cite{Xie2024prb} coincides with our quasi-atomic value of 0.98~$\mu_B$ to have the same field-dependent shift. In a perfectly aligned fully-polarized state, the spin-wave dispersion  is independent of this shift~\cite{Li_Chen2018}. We obtain a very good quantitative agreement with experimental dispersion (Fig.~\ref{fig:CsCeSe2}c) along the $\Gamma-M-Y$ path in the reciprocal space (sketched in Fig.~\ref{fig:CsCeSe2}b) when the full calcualted IEI including the dfIE contribution are used.

 Without dfIE, i.~e., with only superexchange included, the ground-state magnetic order is still $yz$-stripe, but with a large $z$-component of the ordered moment ($|\langle S_z \rangle|$=0.35). The dispersion in the fully-polarized state is strongly underestimated (Fig.~\ref{fig:CsCeSe2}d).  

In order to evaluate magnetic excitations at zero field along the same $\vq$-path, we summed up the contributions  the for three equivalent paths in the reciprocal space shown in Fig.~\ref{fig:CsCeSe2}b to account for equally populated domains of the $yz$-order. The resulting theoretical INS (Fig~\ref{fig:CsCeSe2}e) is in good qualitative agreement with the experimental one   (Fig~\ref{fig:CsCeSe2}f) albeit with the excitation energies underestimated by about 25\% compared to experiment.

\begin{figure}[!b]
	\centering
	\includegraphics[width=1\columnwidth]{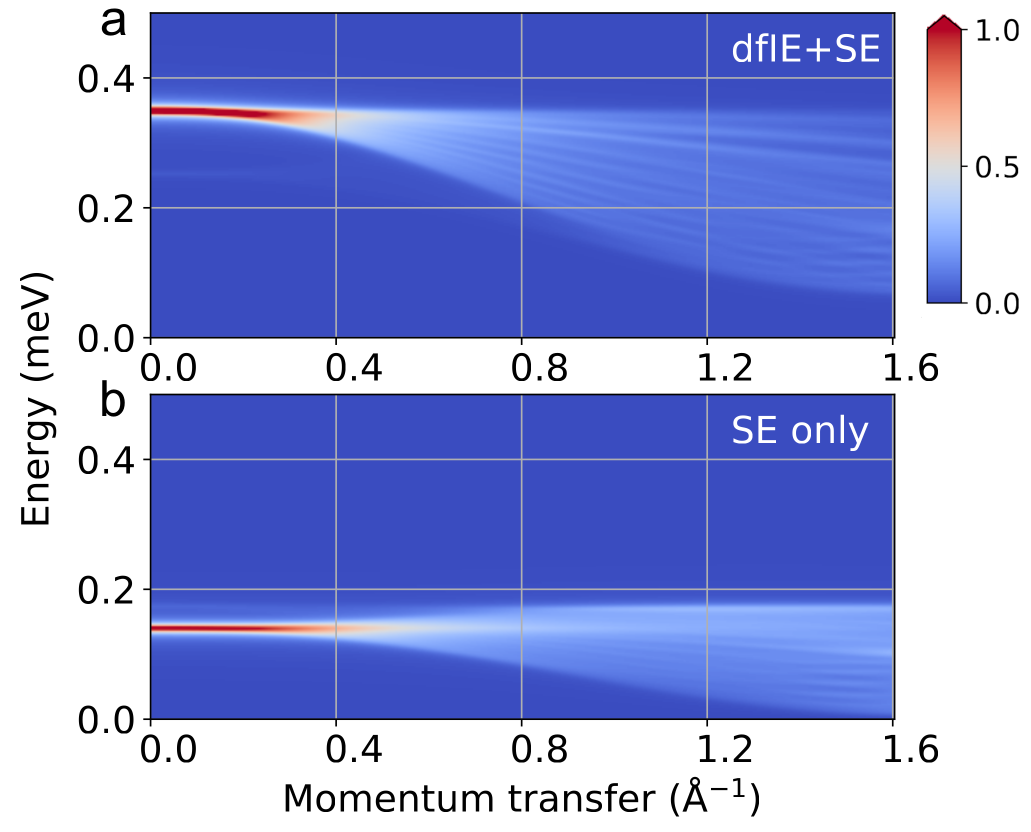}
	\caption{(a). 
	Spherically-averaged INS intensity of KCeS$_2$ calculated using the full IEI Hamiltonian with 5$d$-mediated indirect exchange (dfIE) included (a) and using  superexchange (SE) only (b).}
	\label{fig:KCeS2}
\end{figure}

We apply the same approach to calculate INS intensity in  KCeS$_2$. 
 Solving the full Hamiltonian including both dfIE and superechange within mean-field results, similarly to CsCeSe$_2$, in an $yz$-order with a nearly in-plane  moment direction ($|\langle S^y \rangle|=0.49$ and   $|\langle S^z \rangle|=0.09$).  The calculated spherically averaged INS intensity (Fig.~\ref{fig:KCeS2}a) features a sharp quasi-dispersionless  branch at about 0.35~meV for $q < 0.3$~\r{A}$^{-1}$, which spreads into a broad diffusive feature from 10 to 35~meV at $q > 1.2$~\r{A}$^{-1}$; this picture agrees quite well qualitatively and quantitatively with the experimental low-$T$ INS data~\cite{Avdoshenko2022}. 
Neglecting the dfIE contribution we find in mean-field, analogously to CsCeSe$_2$, a $yz$-order with a large out-of-plane component ($|\langle S_z \rangle|$=0.36). The corresponding INS spectra (Fig.~\ref{fig:KCeS2}b) features the high-intensity branch at an energy that is about half the experimental value.

{\it Outlook.}  The generalized FT-HI method introduced in this Letter is not, per se, limited to insulators and remains applicable as long as the correlated $f$-shell is localized. A wide range of applications is hereby possible, including RE metals and local-moment RE intermetallics,  localized actinide systems, as well as other RE insulating systems, like the  highly frustrated RE pyrochlores. \newver{In the context of RE metals and intermetallics, the IEI include the "virtual mixing interaction" due to 4$f$ electron hopping as well as the contribution generated through the $U_{fd}$ on-site Coulomb repulsion~\cite{Fulde1985,Hirst1978}; both contributions are often called "RKKY" in the literature and both are taken into account by the generalized FT-HI}. The method can be straightforwardly extended to multiple itinerant shells  coupled through an on-site intershell Coullomb interaction to the localized quasi-atomic 4$f$ one,  e.~g.,  both the RE  5$d$ and 6$s$ shell can be included simultaneously. The 4$f$-6$s$  Hund's rule interaction was recently shown to be sizable~\cite{Pivetta2020}.   An important advantage of the method is its ability to separately  evaluate the contributions of various exchange mechanisms, while employing the same framework and the same set of correlated orbitals. 
\newver{The generalized FT-HI method also preserves the principal advantage of the original formulation -- that a complete set of matrix elements is derived for each pair of interacting sites  -- including all allowed multipolar interactions.} 
 The  limitation of its present implementation is that an experimental input is required to define $U_{fd}$ (though the most material sensitive parameter $F^0$ is irrelevant). Combining the generalized FT-HI with an approach for {\it ab initio} determination of the $f-d$ on-site interaction, like, e.~g., the constrained random-phase approximation ~\cite{Aryasetiawan2004}, is, therefore, a  promising avenue for its future development.

{\it Acknowledgments. } The author is grateful to Dario Fiore Mosca for useful discussions and to the CPHT computer team for support.


%

\end{document}


\title{Supplemental Material for '5$d$-mediated indirect exchange and effective spin Hamiltonians in Ce triangular-lattice delafossites'}

\author{Leonid V. Pourovskii}

\affiliation{CPHT, CNRS, \'Ecole polytechnique, Institut Polytechnique de Paris, 91120 Palaiseau, France}
\affiliation{Coll\`ege de France, Université PSL, 11 place Marcelin Berthelot, 75005 Paris, France}

\date{\today}	
\maketitle

\section{Outline of FT-HI approach}
In order to render this paper self-contained, we briefly outline the FT-HI approach~\cite{Pourovskii2016} here. We start by assuming the following form for the low-energy effective quantum Hamiltonian of the system:
\beq\label{eq:Hle}
H_{\mathrm{eff}}=H_{1s}+H_{\mathrm{IEI}}=H_{1s}+\sum_{\langle ij\rangle}\sum_{\alpha\beta}V^{\alpha\beta}_{ij}O^{\alpha}(i)O^{\beta}(j),
\eeq
where $H_{1s}$ is the one-site (typically, crystal field) term, $H_{\mathrm{IEI}}$ is  the two-site IEI one, $O^{\alpha}(i)$ are operators encoding degrees of freedom within the ground-state (GS) multiplet of a given site and labeled by an index $\alpha$. The GS multiplet is  labeled by (pseudo)-angular momentum $J$ and spanned by its $|JM\rangle$ eigenstates, with $M=-J,-J+1,...,J$.  The operators $O^{\alpha}$ are traceless thus conserving the total occupancy within the GS multiplet, but otherwise their form is arbitrary (they can  be angular momentum operators, Stevens operators, spherical tensors etc.). In this section we abbreviate $|JM_1\rangle$ at site $i$ as $|1_i\rangle$. Hereby, two-site matrix elements of (\ref{eq:Hle}) read
\beq\label{eq:MM}
\langle 1_i3_j|H_{\mathrm{IEI}}|2_i4_j\rangle=\sum_{\alpha\beta}V_{ij}^{\alpha\beta}O_{12}^{\alpha}O_{34}^{\beta}.
\eeq
Using the operators $\rho^{M_1M_2}\equiv \rho^{12}$ introduced in the main text, any traceless operator $O^{\alpha}$ can be trivially written as $\sum_{12}O_{12}^{\alpha}\rho^{12}$. Inserting this into $H_{\mathrm{IEI}}$ and using  (\ref{eq:MM}) one finds:
$$
H_{\mathrm{IEI}}=\sum_{\langle ij\rangle}\sum_{1234}\langle 1_i3_j|H_{\mathrm{IEI}}|2_i4_j\rangle\rho_i^{12}\rho_j^{34}.
$$

The mean-field free energy $\Omega_{\textrm{MF}}$ of (\ref{eq:Hle}) for a chosen set of expectation values of all $\rho$ at all sites then   reads~\cite{Georges2004,Plefka1982,Georges1991}:
$$
\Omega_{\mathrm{MF}}=\Omega_{1s}+\sum_{\langle ij\rangle}\sum_{1234}\langle 1_i3_j|H_{\mathrm{IEI}}|2_i4_j\rangle \langle\rho_i^{12}\rangle\langle\rho_j^{34}\rangle,
$$
where the first term comprises single-site entropic and energetic contributions and the second term is the mean-field total energy contribution due to the 2-site term. Therefore the response to  two-site fluctuation of $\rho$ is given by the corresponding matrix element of the two-site Hamiltonian:
$$
\frac{\delta^2\Omega_{\mathrm{MF}}}{\delta \rho_i^{12}\delta \rho_j^{34}}=\langle1_i3_j|H_{\mathrm{IEI}}|2_i 4_j\rangle.
$$

If the low-energy physics of the system is represented by (\ref{eq:Hle}) then  its full electronic model including all energy scales and solved within a (dynamical) mean-field approximation should have the same response, $\frac{\delta^2\Omega_{\mathrm{MF}}}{\delta \rho_i^{12}\delta \rho_j^{34}}=\frac{\delta^2\Omega_{\mathrm{DFT+DMFT}}}{\delta \rho_i^{12}\delta \rho_j^{34}}$, from which eq.~3 of the main text follows.

In order to evaluate $\frac{\delta^2\Omega_{\mathrm{DFT+DMFT}}}{\delta \rho_i^{12}\delta \rho_j^{34}}$, Ref.~\cite{Pourovskii2016} employs a force theorem approach evaluating the response 
of the kinetic energy term in $\Omega_{\mathrm{DFT+DMFT}}$ only. The latter is given by 

$$\frac{1}{\beta} \mathrm{Tr}\, \mathrm{ln} \left[G_{\mathrm{latt}}^{\vk,n}\right]=-\frac{1}{\beta} \mathrm{Tr}\,  \mathrm{ln}\left[i\omega_n+\mu-H_{\mathrm{KS}} -P_{\vk}^{\dagger}(\Sigma(i\omega_n)-\Sigma_{\mathrm{DC}})P_{\vk}\right],$$ 
where $\omega_n$ is the fermionic Matsubara frequency, $\mu$ is the chemical potential,  $H_{\mathrm{KS}}$ is the Kohn-Sham Hamiltonian, $\Sigma_{\mathrm{DC}}$ is the self-energy double counting term, other quantities are defined in the main text. Within the force-theorem approximation, only the same-site self-energy $\Sigma_i(i\omega_n)$ is affected by a fluctuation in $\rho_i$ leading to the FT-HI equation, eq.~5 of the main text, for   $\langle 1_i3_j|H_{\mathrm{IEI}}|2_i 4_j\rangle$.

\section{DFT+HI calculations} We calculated all three compounds in DFT+HI using their experimental lattice structures~\cite{Xie2024prb,Kulbakov2021,Ortiz2022} employing the Wien2k linearized augmented-plane-wave (LAPW) code~\cite{Wien2k}, TRIQS library implementation of DFT+DMFT~\cite{Parcollet2015,Aichhorn2016} and the Hubbard-I implementation provided by the MagInt package~\cite{magint}. 
In those DFT+HI calculations we employ 1000 $\mathbf{k}$-points in the full Brillouin zone, the LAPW basis cutoff $R_{\mathrm{MIN}}K_{\mathrm{MAX}}$=8, the spin-orbit coupling is included through the standard second-variation approach, the LDA is used as the DFT exchange-correlation potential. The double counting correction is calculated in accordance with the fully-localized-limit formula using the nominal occupancy $f^1$. 

The projected Wannier correlated bases~\cite{Aichhorn2009} were constructed using the same "large" ([-9.5:9.5]~eV, for $X$-p and Ce-5$d$)   and the "small" ([-1.63:1.56]~eV for Ce-4$f$) energy windows for all 3 compounds, the latter window chosen such as to include 4$f$ only excluding the rest in order to properly account for hybridization contribution to the crystal field splitting~\cite{Pourovskii2020}. The resulting  full set of  Ce 5$d$, Ce 4$f$, and $X$ $p$ correlated orbitals is then orthonormalized using the standard prescription of the projective construction approach~\cite{Aichhorn2009} to obtain a proper Wannier basis.  

We specified  the rotationally invariant $ff$ Coulomb vertex using the standard values  $F^0=U=6$~eV and $J_H=0.7$~eV for Ce$^{3+}$  that were also used in Ref.~\cite{Pourovskii2025_with_Alex}. The $fd$ rotationally invariant Coulomb vertex was defined~\cite{Cowan_book}  using the Slater integrals values $F^2=2.82$, $F^4=1.40$, $G^1=1.20$, $G^3=1.04$, and $G^5=0.81$~eV  extracted from optical measurements of $4f^2\to 4f^15d^1$ excitations in Pr$^{3+}$ doped in LiYF$_4$ host~\cite{Burdick2007}. The use of  Pr$^{3+}$ $fd$ Slater parameters for Ce$^{3+}$ is justified by those parameters being basically constant  between Pr$^{3+}$ and Nd$^{3+}$ with the change  $\lesssim$1\%~\cite{Burdick2007}. We note that the  $fd$ Slater parameter  $F^0$, which is expected to be strongly system dependent, does not contribute to the HF self-energy response (eq. 4 of the main text) since  the fluctuations $\delta \rho^{M_1M_2}$ conserve the 4$f$ shell occupancy.

\begin{table}[!b]
			\centering
			\renewcommand{\arraystretch}{1.5}
			\begin{tabular}{ c | l }             				
				\textbf{$E$ (meV)\hspace{2mm}}& \hspace{2mm} \textbf{Wavefunction} \\
				\hline 
				0 & \hspace{2mm} $0.884|5/2;\mp 1/2\rangle\pm 0.466|5/2;\pm 5/2\rangle-0.037|7/2;\mp 7/2\rangle$   \\
				38 & \hspace{2mm} $0.995|5/2;\pm 3/2\rangle\mp 0.073|5/2;\mp 3/2\rangle\mp0.060|7/2;\pm 3/2\rangle +0.040|7/2;\mp 3/2\rangle$    \\
				60  &  \hspace{2mm} $0.881|5/2;\pm 5/2\rangle\mp 0.463|5/2;\mp 1/2\rangle\pm 0.068|7/2;\mp 7/2\rangle+0.067|7/2;\mp 1/2\rangle$  \\
	\end{tabular}
\captionsetup{justification=centering}
\caption{Calculated CF energies $E$ and CF Kramers doublet wavefunctions in CsCeSe$_2$.}
\label{tab:CF_WF}
\end{table}

Our self-consistent DFT+HI calculations  were carried out employing the averaging over the  $^2F_{5/2}$ GSM  of Ce 4$f^1$ to suppress the unphysical DFT self-interaction contribution to crystal field (CF)~\cite{Delange2017}. 
The calculated CF splitting and GS doublet agree rather well  with experimental estimates~\cite{Xie2024prb,Kulbakov2021,Ortiz2022}. However, as in Ref.~\cite{Pourovskii2025_with_Alex}, we do not reproduce the "superfluous" third CF mode experimentally observed in all three systems. This mode cannot be explained within a single-ion CF model, see, e.~g., Refs.~\cite{Ortiz2022,Kulbakov2021,Xie2024prb}. In CsCeSe$_2$, we  find a strong in-plane anisotropy of the GS $g-$tensor with in-plane $g_{x}=g_y$=1.96 and  out-of plane $g_z$=-0.20 agreeing with the range of values obtained by different experimental probes \cite{Xie2024prb,Xie2024prl}, see  Table~\ref{tab:CF_WF} for the CF splitting and eigenstates in CsCeSe$_2$.  The calculated CF splittings in KCeS$_2$ and RbCeO$_2$ are the same as reported in Ref.~\cite{Pourovskii2025_with_Alex} apart from minute differences due to slight changes in the 4$f$ correlated basis.

\section{Intersite exchange interactions}

The IEI interaction $\hat{J}_{ij}$ in the spin-1/2 Hamiltonian (eq.~6 of the main text) is obtained from the FT-HI–calculated matrix elements $\langle M^i_1M_3^j|H_{\mathrm{IEI}}|M^i_2M^j_4\rangle$ as follows: 
$$
J_{ij}^{pp'}=2\langle M^i_1M_3^j|H_{\mathrm{IEI}}|M^i_2M^j_4\rangle S^{p}_{M_2M_1}S^{p'}_{M_4M_3},
$$
where $S^{p(p')}$ are the spin-1/2 operators,  $p(p')=x,y,z$,  the prefactor 2 stems from the spin-1/2 operators' normalization.

In Table~\ref{tab:Js} we list the calculated full IEI interactions, with indirect exchange included, for the NN bond along the $x$ axis and the NNN bond along the $y$ axis, see Fig.2a. Apart from the NN anisotropy, we include also the full NNN anisotropy. To distinguish the NN and NNN parameters, the latter are primed ($J'$, $\Delta'$ etc.). We also list the average value,   $J^{\mathrm{int}}=Tr[\hat{J}^{\mathrm{int}}]/3$, for the shortest interlayer coupling.

\begin{table}[!bt]
	\begin{center}
		\begin{ruledtabular}
			\renewcommand{\arraystretch}{1.2}
			\begin{tabular}{l c c c}
				& CsCeSe$_2$ & KCeS$_2$ & RbCeO$_2$  \\
				\hline
				$J$ (meV) &   0.049  &     0.055   &   0.299 \\
				$\Delta$ & 0.574  &    0.971    &   1.110 \\
				$J_{\pm\pm}/J$ &  0.572  &    0.717   &   0.133 \\
				$|J_{z\pm}/J|$  & 0.138   &   0.254    &  0.040 \\
				\hline
				$J^{\prime}/J$&  -0.013  &   -0.075  &   -0.069 \\
				$\Delta^{\prime}$  & 3.548   &   0.555   &   0.640 \\
				$J_{\pm\pm}^{\prime}/J^{\prime}$ &-1.710   &  -0.207  &   -0.018 \\
				
				$|J_{z\pm}^{\prime}/J^{\prime}|$ & 1.517   &  0.181  &    0.115 \\
				\hline
				$J^{^{\mathrm{int}}}/J$ & -0.008  &   -0.027   &  -0.036 \\
			\end{tabular}
		\end{ruledtabular}
	\end{center}
	\caption{Calculated IEI with indirect exchange included.}
	\label{tab:Js} 
\end{table}

\section{INS intensity calculations within RPA}

The INS intensity at the energy  $E$ and momentum $\vq$ reads
$$
I(\vq,E)=F^2(|\vq|)\sum_{\substack{pp' \\ ii'}}\left(\delta_{pp'}-\frac{q_{p}q_{p'}}{q^2}\right)g_pg_{p'}\Im \chi^{ii'}_{pp'}(\vq,E),
$$ 
where $F(|\vq|)$ is the Ce$^{3+}$ form factor in the dipole approximation. $\chi(\vq,E)$ is the dynamical susceptibility, $p$ and $p'=x,y,z$, $i$ and $i'$ run over the sites of the magnetic unit cell. We evaluate the dynamical susceptibility $\chi(\vq,E)$ using the usual random-phase approximation formula (for a general introduction into the random-phase approximation and its application to magnetic excitations see, e.~g., Ref.~\cite{Jensen1991}) that reads
$$
\bar{\chi}^{-1}(\vq,E)=\bar{\chi}^{-1}_0(E)+\bar{J}_{\vq},
$$
where $\chi_0(E)$ is the single-site susceptibility evaluated from the on-site mean-field  potential in the ordered phase, $J_{\vq}$ is the   Fourier-transformed calculated IEI,  the  bar $\bar{ }$ labels matrices in the $p$ and $i$ indices. The Ce$^{3+}$  form-factor $F(|\vq|)$ was calculated in accordance with the standard formula employing the radial integrals $\langle j_0 (q)\rangle$ and  $\langle j_2(q) \rangle$ obtained  in Ref.~\cite{Lovesey2020} using a relativistic Hartree-Fock approach~\cite{Cowan_book}. 

In order to simulate polycrystalline INS data in KCeS$_2$ we 
average, for each $q=|\vq|$, the intensity $I(\vq,E)$
as $I(q,E)=\sum_{\vq,|\vq|=q}I(\vq,E)$ 
over an equidistant spherical $\vq$-mesh containing 642 $\vq$-points.

%